\documentclass[fleqn,10pt]{wlscirep}

\usepackage{siunitx} 
\usepackage{enumitem} 
\usepackage{graphicx}
\usepackage{epstopdf}
\usepackage{cite}

\title{Population Fluctuation Promotes Cooperation in Networks}

\author[1,*]{Steve Miller}
\author[1]{Joshua Knowles}
\affil[1]{University of Manchester, School of Computer Science - Machine Learning \& Optimisation Group, UK}

\affil[*]{stevemiller.gm@gmail.com}


\begin{abstract}
We consider the problem of explaining the emergence and evolution of cooperation in dynamic network-structured populations. Building on seminal work by Poncela et al., which shows how cooperation (in one-shot prisoner’s dilemma) is supported in growing populations by an evolutionary preferential attachment (EPA) model, we investigate the effect of fluctuations in the population size. We find that a fluctuating model -- based on repeated population growth and truncation -- is more robust than Poncela et al.'s in that cooperation flourishes for a wider variety of initial conditions. In terms of both the temptation to defect, and the types of strategies present in the founder network, the fluctuating population is found to lead more securely to cooperation. Further, we find that this model will also support the emergence of cooperation from pre-existing non-cooperative random networks. This model, like Poncela et al.’s, does not require agents to have memory, recognition of other agents, or other cognitive abilities, and so may suggest a more general explanation of the emergence of cooperation in early evolutionary transitions, than mechanisms such as kin selection, direct and indirect reciprocity. 
\end{abstract}

\begin{document}

\flushbottom
\maketitle
\thispagestyle{empty}

\section*{Introduction}
Cooperation among organisms is observed, both within and between species, throughout the natural world. It is necessary for the organization and functioning of societies, from insect to human. Cooperation is also posited as essential to the evolutionary development of complex forms of life from simpler ones, such as the evolutionary transition from prokaryotes to eukaryotes or the development of multicellular organisms~\cite{smith_major_1997}{}. A widespread phenomenon in nature, cooperative behaviour has been studied in detail in a wide variety of situations and lifeforms; in viruses~\cite{shirogane_cooperation:_2013}{}, bacteria~\cite{crespi_evolution_2001}{}, insects\cite{wilson_insect_1971}{}, fish\cite{milinski_tit_1987}{}, birds\cite{stacey_cooperative_1990}{}, mammals\cite{clutton-brock_cooperation_2001}{}, primates\cite{mendres_capuchins_2000} and of course in humans\cite{axelrod_evolution_1981}{}, where the evolution of cooperation has been linked to the development of language\cite{nowak_evolution_1999}{}.

The riddle of cooperation is how to resolve the tension between the ubiquitous existence of cooperation in the natural world, and the competitive struggle for survival between organisms (or genes or groups), that is an essential ingredient of the Darwinian evolutionary perspective. Based on existing theories\cite{hamilton_genetical_1964, trivers_evolution_1971, grafen_natural_1984}{}, Nowak\cite{nowak_five_2006} describes a framework of enabling mechanisms to address the existence of cooperation under a range of differing scenarios. This framework consists of the following five mechanisms:  \textit{kin selection, direct} and \textit{indirect reciprocity, multi-level selection} and \textit{network reciprocity}{}.  These mechanisms have been developed and much studied within the flourishing area of evolutionary game theory, and to a lesser extent in the simulated evolution and artificial life areas (in computer science).

Our interest in this paper is \textit{network reciprocity}, where the interactions between organisms in relation to their network structure, offer an explanation of cooperation. This mechanism is important for two reasons. First, whilst cooperation is widespread and found in a broad range of scenarios in the real world, many of the mechanisms that have been proposed to explain it require specific conditions such as familial relationships (for kin selection), the ability to recognise or remember (for direct and indirect reciprocity) and transient competing groups (for multi-level selection) (see Nowak\cite{nowak_five_2006} for specific details). The requirement for such conditions limits the use of each of these mechanisms as a more general explanation for a widespread phenomenon. Secondly, the more specific behavioural or cognitive abilities required by some of these mechanisms precludes their use in explaining the role of cooperation in early evolutionary transitions. Network-based mechanisms which focus on simple agents potentially offer explanations which do not require such abilities. All forms of life exist in some sort of relationship with other individuals, or environments, and as a result can all be considered to exist within networks. Thus the network explanation has ingredients for generality which are lacking in the other models. 

It has been shown that networks having heterogeneous structure can demonstrate cooperation in situations where individuals interact with differing degrees\cite{santos_new_2006}{}, effectively by \textit{clustering}. Populations studied in this way are represented in the form of a network, with individuals existing at the nodes of the network and connections represented by edges.  The degree of an individual node indicates the number of neighbour nodes it is connected to. Heterogeneity refers to the range of the degree distribution within the network. Behaviour in these networks can be investigated using the prisoner’s dilemma game which is widely adopted as a metaphor for cooperation. 

The majority of studies\cite{santos_new_2006,nowak_evolutionary_1992,szabo_evolutionary_2007} investigating cooperation with regards to network structure have focused on static networks and hence consider the behaviour of agents distinct from the networks within which they exist. Specifically in these works, the behaviour of the individuals within the network has no effect on their environment. More recently however, the interaction between behaviour and the development of network structure has been considered in an interesting development which shows promise in understanding evolutionary origins of cooperation. The Evolutionary Preferential Attachment (EPA) model developed by Poncela et al.\cite{poncela_complex_2008} proposes a fitness-based coevolutionary mechanism where scale-free networks, which are supportive of cooperation, emerge in a way that is influenced by the behaviour of agents connecting to the network. 

There is a large body of work devoted to coevolutionary investigations of cooperation (see Perc and Szolnoki\cite{perc_coevolutionary_2010} for a review) of which a subset focus on coevolutionary studies within networks\cite{ebel_coevolutionary_2002, pacheco_coevolution_2006, szolnoki_making_2008, cardillo_co-evolution_2010 }{}.  However the EPA approach of Poncela, which we investigate further in this report, is notable in that it addresses an area that seems to have received very little attention: Specifically it explores how environment affects the behaviour of individuals, \emph{simultaneously} with how such individuals, in return, affect their developing environment.  

In the EPA model, new network nodes connect preferentially to existing network nodes that have higher fitness scores. Accumulated fitness scores arise from agents located on nodes within the network playing one-shot prisoner’s dilemma with their neighbours. Strategies are subsequently updated on a probabilistic basis by comparison with the relative fitness of a randomly-selected neighbour. The linking of evolutionary agent behaviours to their environment in this way has been shown to promote cooperation, whilst the use of fitness rather than degree allows for a broader and more natural representation of preferential attachment\cite{nguyen_fitness-based_2012}{}. Whilst further exploring the role of scale-free network growth with regards to cooperation (which of itself is interesting) the EPA model also implements one-shot rather than iterated prisoner's dilemma and it utilises agents having unconditional (fixed) strategies; hence it potentially presents a very simple minimal model, in the light of reported findings, for the coemergence of cooperation and scale-free networks. 

Our investigations here are driven by two observations regarding the EPA model. First, we note that the model achieves a fixed network structure very early within simulations, from which point onwards agent behaviour has no effect on network structure. Secondly, whilst the EPA model supports the growth of cooperation in networks from a founder population consisting solely of cooperators, it does not achieve similar levels of success for networks grown from defectors. The broader question of how cooperation emerges requires an answer which can generalise to explain emergence from populations that are not assumed cooperative initially; hence, in this work we investigate networks grown from founder populations which may be cooperators \emph{or} defectors. 

We introduce a modification to the EPA model (see Methods for details) which we consider an abstraction common to most, if not all, real populations, that of population size fluctuation. We investigate whether the resulting opportunity for the agents to continually modify the network, leads to increased levels of cooperation in the population. We achieve this fluctuation by truncating the evolved network whenever it reaches the specified maximum size. At truncation, agents are selected for deletion on the basis of fitness. Those least fit are removed from the network. The network is grown and truncated repeatedly until the simulation is ceased. The original EPA model offered a limited period of time for agents to initially affect the structure of their network. Our modification makes this `window of opportunity' repeatedly available. Whilst a small number of interesting studies have explored the effect on cooperation of deleting network links\cite{zimmermann_cooperation_2005, santos_cooperation_2006, pacheco_active_2006, traulsen_evolutionary_2009}{}, or to a much lesser extent, nodes\cite{ perc_evolution_2009, szolnoki_impact_2009, ichinose_robustness_2013}{}, the process we have implemented here differs in that it specifically targets individuals (nodes) on the basis of least-fitness.  As such it has a very clear analogue in nature, in terms of natural selection. 

The question, ``How does cooperation emerge?'' can be considered from two extreme perspectives, firstly the scenario where cooperation develops within a population from its very earliest origins and secondly in terms of its emergence within a pre-existing non-cooperative network. In reality cooperation may occur anywhere within a spectrum bounded by these two extrema, at different times for different sets of events and circumstances, therefore a network-based mechanism to explain this phenomenon should be able to deal with either extreme and other positions in between. In testing this model, we investigate scenarios where cooperation may develop as a population grows from its founder members and we also apply the model to pre-existing randomly structured networks.

\section*{Results}
Unless stated otherwise in the text, the general outline of the evolutionary process by which our results were obtained, occurs for one generation as follows:
\begin{enumerate}[noitemsep]
	\item  \textit{Play prisoner's dilemma}: Each agent plays one-shot prisoner's dilemma with all neighbours and achieves a fitness score that is the sum of all the payoffs.
	\item \textit{Update strategies}: Those strategies that result in low scores are replaced on a probabilistic basis by comparison with the strategies of randomly selected neighbours.
	\item \textit{Grow network}: A specified number (we used 10 in all our simulations) of new nodes are added to the network, connecting to $m$ distinct existing nodes via \textit{m} edges using either EPA or CRA. 
	\item \textit{Remove nodes (only in the case of attrition models)}: If the network has reached maximum size, it is pruned by a truncation process that removes agents on the basis of least fitness.
\end{enumerate}

\noindent Full details on the specifics of the implementations are provided in the methods section.

\subsection*{Results for networks grown from founder populations}
We investigate the effect of population fluctuation on networks grown from founder populations consisting of three nodes. \\

\begin{figure}[!b]
	\centering
	\includegraphics[width=17cm]{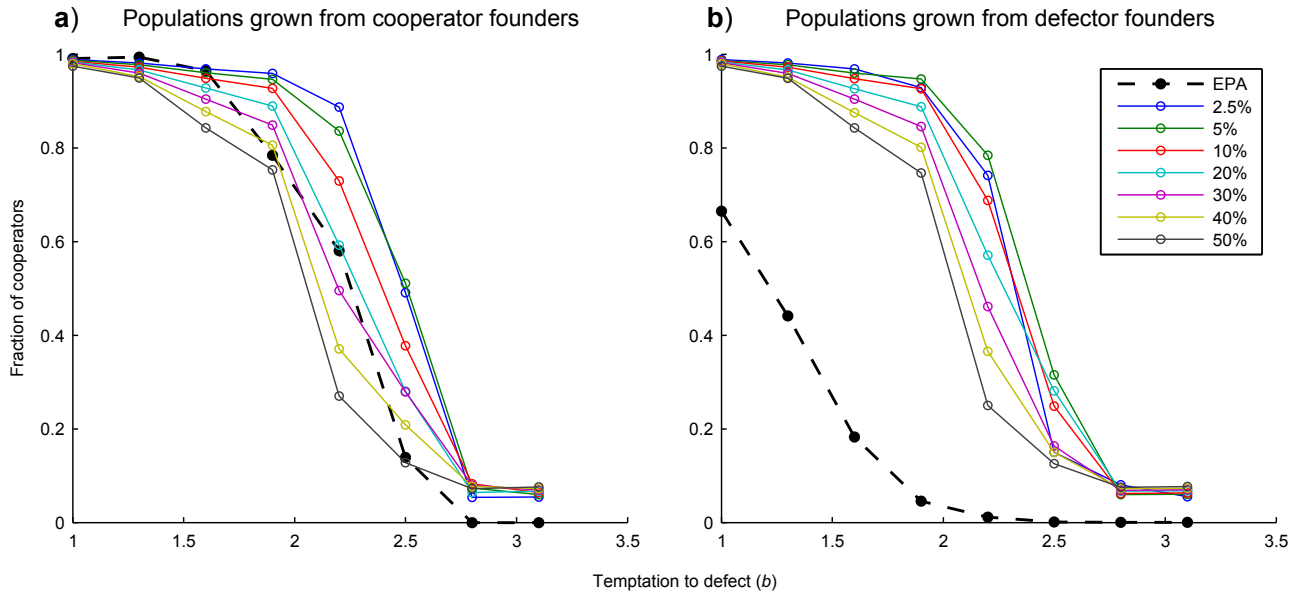}
	\caption{Effect of truncation size on cooperation. Simulations were run for 2000 generations using the EPA and fluctuation models at a range of $b$  values. The graphs plot the fraction of cooperators in the population against  $b$  values. Each point on the graph  represents  an  average of 25 individual results. Each of the individual results is in turn an average of the last 20 generations of a simulation replicate. Plots are shown from fluctuation model simulations using 2.5 to 50\% network  truncation. An EPA simulation is also shown which does not feature any truncation.  (\textbf{a})  illustrates results grown from founder populations of 3 cooperators. (\textbf{b}) illustrates results grown from founder populations of 3 defectors. See Figures \ref{fig:Example_simulation_plots_C_founders} and \ref{fig:Example_simulation_plots_D_founders}, and accompanying text for more detailed discussion of within-simulation data variability.}
	\label{fig:Effect_of_truncation}
\end{figure}

\noindent\textbf{Low levels of truncation result in increased levels of cooperation.} For simulations starting from founder networks consisting solely of cooperators, we achieved similar profiles to those from the EPA model, however when lower levels of truncation (less than 20\%) were used we were able to demonstrate consistently higher levels of cooperation than the EPA model for values of $b$ (the temptation to defect) greater than 1.6 (see Figure\ref{fig:Effect_of_truncation}a). Highest levels of cooperation were achieved using as little as 2.5\% and 5\% truncation. We observed that cooperation does not reduce to the levels seen for EPA until truncation values are reduced to as little as 0.1\% (not shown).  Whilst large percentage truncations risk deleting some of the higher fitness nodes which give the network its heterogeneous (power-law) degree distribution and hence aid cooperation, small truncation percentages will be focused on low-fitness, low-connectivity nodes, the deletion of which is unlikely to have such a detrimental effect.  Also, given small truncation values, truncation events will occur at higher frequencies, thus supplying a steady `drip-feed' of new nodes which will attach to existing hubs by preferential attachment and hence continually promote a power-law degree distribution within the network.   

The reason that the EPA model can achieve higher levels of cooperation at $b$ = 1 to 1.6 than the fluctuation model is because whilst it is possible for the EPA model to be completely overrun by cooperators, in the fluctuation model however, repeated truncation prevents such a situation occurring.  Defectors are being added to the population after every truncation event. A similar constraint also applies at the other end of the scale, with regards to very low levels of cooperation. So there are limits below 1 and above 0 which are a result of the truncation size and frequency. These limits restrict the range of cooperation values achievable by the fluctuation model. \\

\noindent\textbf{Cooperation occurs even for populations that are founded entirely with defectors.} Our results starting from founder populations consisting solely of defectors show an increase compared with levels of cooperation achieved by the EPA model (see Figure \ref{fig:Effect_of_truncation}b). Further, we note that final levels of cooperation arising from the fluctuation model for networks founded from cooperator and from defector strategies were almost indistinguishable statistically: we tested the dependence of the final cooperation levels observed as a function of $b$, the temptation to defect, and the founding strategy type (C or D), using a nonparametric Sign Test\cite{conover_practical_1999} (see Table \ref{tab:Sign_Test}). \\

\begin{table}[h]
\begin{center}
\begin{tabular}{l|rrS[table-format=1.4]l}
Model & $n$ & $k$ & \multicolumn{1}{c}{$p$-value} &\\ \hline \\[-2mm]
EPA & 178 & 151 & {$<$ 2.20} & \hspace{-4mm}e$-16$\\
T 2.5\% & 240 & 143 & 0.003587 &\\
T 5\% & 240 & 129 & 0.2724 &\\
T 10\% & 240 & 126 & 0.4778 &\\
T 20\% & 240 & 137 & 0.03294 &\\
T 30\% & 240 & 131 & 0.1751 &\\
T 40\%  & 239 & 136 & 0.03820 & \\
T 50\% & 239 & 129 & 0.2442 &\\ \hline
\end{tabular}
\caption{Results of a nonparametric Sign Test (using a two-tailed exact binomial calculation), comparing the final level of cooperation observed in networks founded with cooperators and networks founded with defectors. For each level of truncation, the $240 * 2$ independent samples were paired by the value of $b$, the temptation to defect. The column $n$ is the number of non-tied sample pairs. The column $k$ is the number of times the C-founded population had a larger final cooperation level than the D-founded population. With the standard EPA model there is strong statistical evidence that the cooperator- and defector-founded networks differ. For the fluctuation model, the evidence is much less clear. Given the power of the test is high here due to the relatively large number of samples used, we can tentatively conclude that there is little or no effect of the type of network founding strategy (cooperator or defector) in those fluctuation models having above 2.5\% truncation.}
\label{tab:Sign_Test}
\end{center}
\end{table}

\begin{figure}[!]
	\centering
	\includegraphics[width=17cm]{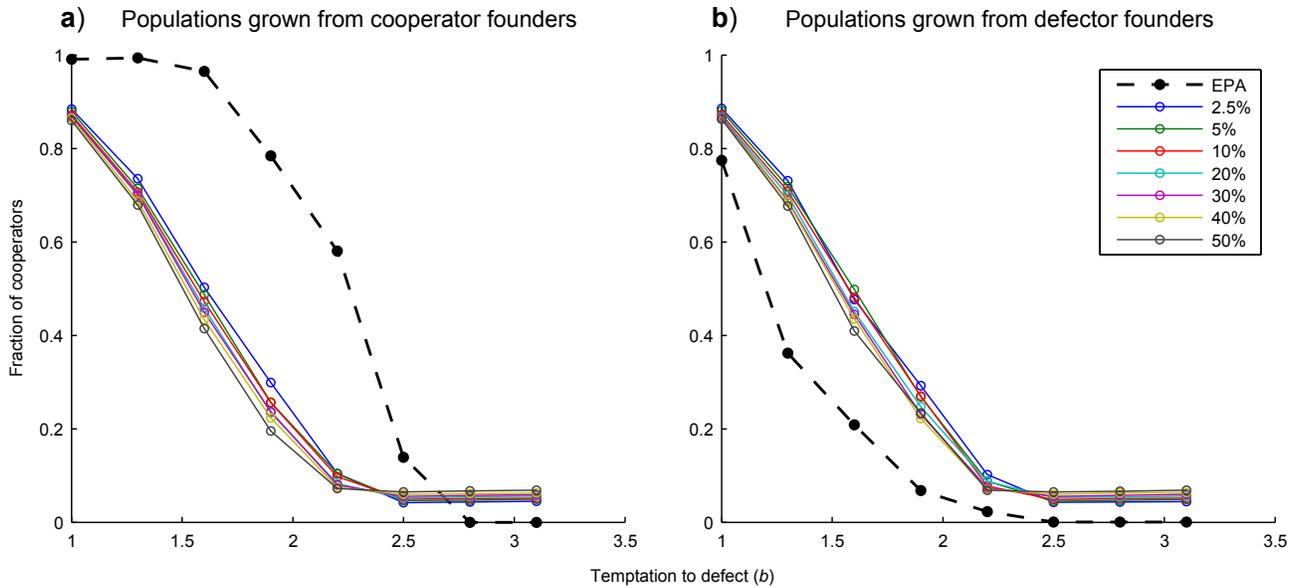}
	\caption{Effect of random rather than weighted selection of nodes for network truncation. Simulations were run as described for Figure \ref{fig:Effect_of_truncation} however, in the fluctuation model, least-fitness based node deletion was replaced with random deletion. Plots are shown from fluctuation model simulations using 2.5 to 50\% network truncation.  For reference, an EPA simulation is also shown which does not feature any truncation. (\textbf{a}) illustrates results grown from founder populations of 3 cooperators. (\textbf{b}) illustrates results grown from founder populations of 3 defectors.}
	\label{fig:Effect_of_random_selection_for_truncation}
\end{figure}

\noindent\textbf{Fluctuation using random selection can still improve cooperation for defector-founded populations.} As a control to the effect seen in the fluctuation model, we repeated the above simulations, deleting nodes randomly rather than on the basis of lowest fitness. Results are illustrated in Figure \ref{fig:Effect_of_random_selection_for_truncation}. By comparing with Figure \ref{fig:Effect_of_truncation}, it can be seen that there is a clear difference in outcomes.  First, for random deletion (Figure \ref{fig:Effect_of_random_selection_for_truncation}), fractions of cooperators present are reduced compared to least fitness (Figure \ref{fig:Effect_of_truncation}); although, we note that levels of cooperation achieved are still independent of the founder population strategy (Figures \ref{fig:Effect_of_random_selection_for_truncation}a and \ref{fig:Effect_of_random_selection_for_truncation}b are approximately equivalent for fluctuation model simulations). Secondly, the percentage truncation parameter no longer appears to have any effect on cooperation (all truncation graphs in Figure \ref{fig:Effect_of_random_selection_for_truncation} look approximately equivalent regardless of \% values).  

Focusing solely on Figure \ref{fig:Effect_of_random_selection_for_truncation}, we now consider the fluctuation model compared to the EPA model.  In the case of networks grown from cooperator-founders (Figure \ref{fig:Effect_of_random_selection_for_truncation}a), EPA demonstrates higher levels of cooperation than the fluctuation simulations.  Truncating the network by a method that simply deletes nodes at random is unsurprisingly, less effective at promoting cooperation than the EPA model which has been shown to be effective for cooperator-founded networks.  In the case of networks grown from defector-founders (Figure \ref{fig:Effect_of_random_selection_for_truncation}b), the fluctuation model still achieves the same results as it did for cooperator-founders (in Figure \ref{fig:Effect_of_random_selection_for_truncation}a).  However in this case, EPA achieves lower levels of performance than the fluctuation model featuring `random' truncation. As we have mentioned EPA is generally less effective at promoting cooperation when networks are grown from defector-founders.

How is the fluctuation model, with random truncation, still able to promote cooperation (albeit at reduced levels compared to targeted truncation)? The random deletion process will inevitably disrupt the formation of the heterogeneous network by deleting the higher degree nodes that are key to the scale-free structure. However, this disruption will be countered by the preferential process for \emph{addition} of replacement nodes which is still fitness-based, i.e. new nodes added will still be preferentially attached to existing nodes of higher fitness. Heterogeneity of degree, (which supports cooperation) still arises, but not to the extent seen for the fluctuation model, which targets least fit nodes for deletion. The preferential attachment process explains why the fluctuation model is still able to have a positive (but reduced) effect on levels of cooperation, even when nodes are deleted randomly.  \\

\begin{figure}[!]
	\centering
	\includegraphics[width=17cm]{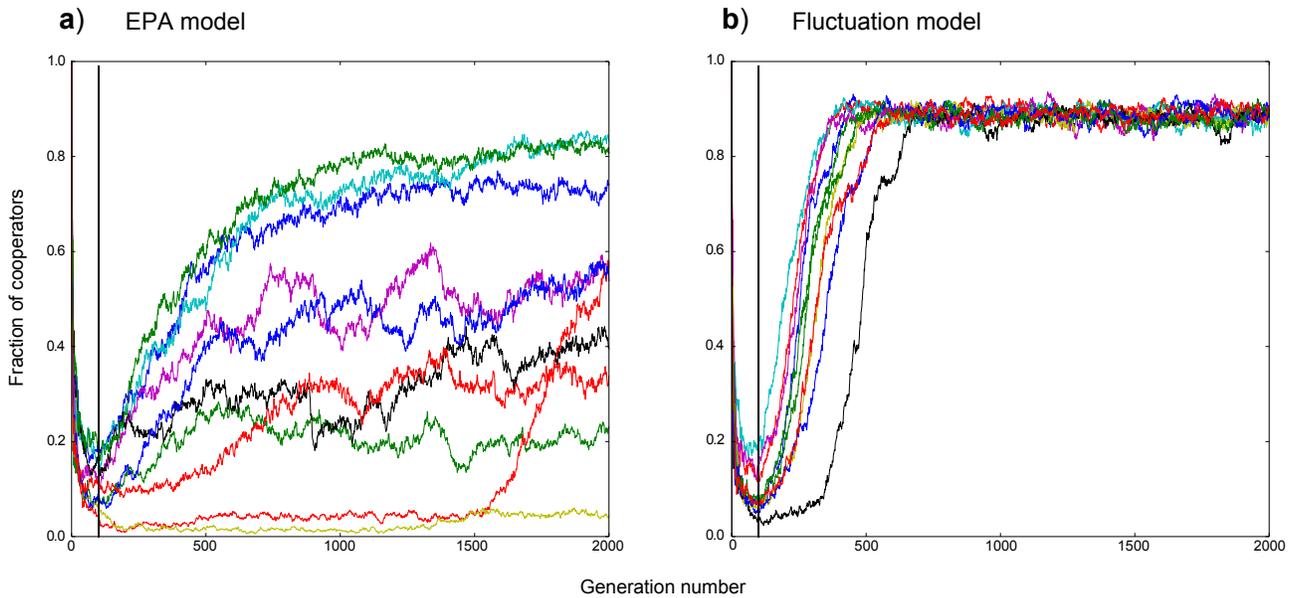}
	\caption{Example simulation plots for EPA and fluctuation models starting from cooperator founders. Plots show the individual time profiles for 10 replicates using a $b$ value of 2.2. (\textbf{a}) shows the EPA model. (\textbf{b}) shows the fluctuation model operating with 2.5\% truncation. Generation 100 is marked in both figures by a vertical black line. This is the point at which the EPA model reaches a fixed network structure, after which no further nodes are added.}
	\label{fig:Example_simulation_plots_C_founders}
\end{figure}

\noindent\textbf{Fluctuation in population size reduces variability within simulation results and increases cooperation.} In Figure \ref{fig:Example_simulation_plots_C_founders}, we provide sample illustrations of time profiles from the EPA and fluctuation models respectively (starting from cooperator founders, $b = 2.2$). The EPA model (Figure \ref{fig:Example_simulation_plots_C_founders}a) results in high variability between different simulation replicates with some replicates being overrun by defectors (i.e. fraction of cooperators $\approx 0$). The fluctuation model (Figure \ref{fig:Example_simulation_plots_C_founders}b) demonstrates far less variability, with clear transitions between two states. Whilst the time at which transition occurs varies, most replicates achieved transitions to a consistent level of cooperation – equivalent to, or greater than the highest level observed from amongst all simulations in a comparable EPA model. 

We considered the possibility that the EPA model may simply require longer for convergence and hence ran extended simulations up to 200,000 generations (not shown).  We did not see any consistent convergence over later generations: whilst some replicates achieved higher levels of cooperation beyond 2,000 generations, others did not, and some oscillated continually. \\ 

\begin{figure}[!]
	\centering
	\includegraphics[width=17cm]{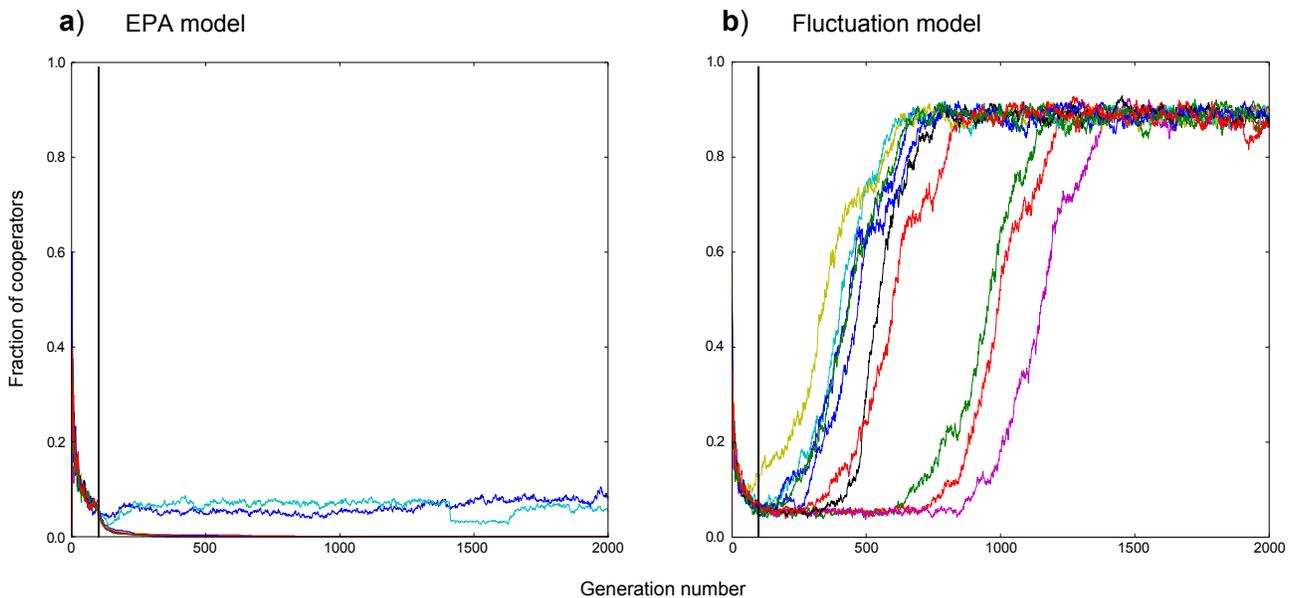}
	\caption{Example simulation plots for EPA and fluctuation models starting from defector founders. Plots show the individual time profiles for 10 replicates using a $b$ value of 2.2. (\textbf{a}) shows the EPA model with network fixation occurring at generation 100. (\textbf{b}) shows the fluctuation model operating with 2.5\% truncation. Generation 100 is marked in both figures by a vertical black line. This is the point at which the EPA model reaches a fixed network structure, after which no further nodes are added.  }
	\label{fig:Example_simulation_plots_D_founders}
\end{figure}

\noindent\textbf{Fluctuation results in dramatic increases in cooperation for networks grown from defectors.} Figure \ref{fig:Example_simulation_plots_D_founders} shows replicate simulations grown from defector founders, using the EPA and fluctuation models respectively. In the EPA model (Figure \ref{fig:Example_simulation_plots_D_founders}a), all replicates are overrun by defectors. In the fluctuation model (Figure \ref{fig:Example_simulation_plots_D_founders}b), all replicates transition to cooperation. 

Ultimately, levels of cooperation achieved are similar for the fluctuation model regardless of whether the founder network is cooperators or defectors. We have however noticed that whilst final outcomes are typically similar for both types of founding strategy, defector-founded simulations tend to result in later times to transitions and greater variation in such times (Figures \ref{fig:Example_simulation_plots_C_founders}b and \ref{fig:Example_simulation_plots_D_founders}b illustrate this). Generally, for cooperator-founded populations, with $b$ values where cooperation was able to arise, we observed transition of the majority ($> 95\%$) of replicates within our typical simulation period of 2000 generations, with delayed transitions becoming more common given increasing $b$ values.  For defector-founded populations, delayed transitions occurred more frequently and to achieve consistent results ($> 95\%$ transitioned) required 20,000 generations.  

\begin{figure}[!]
	\centering
	\includegraphics[width=17cm]{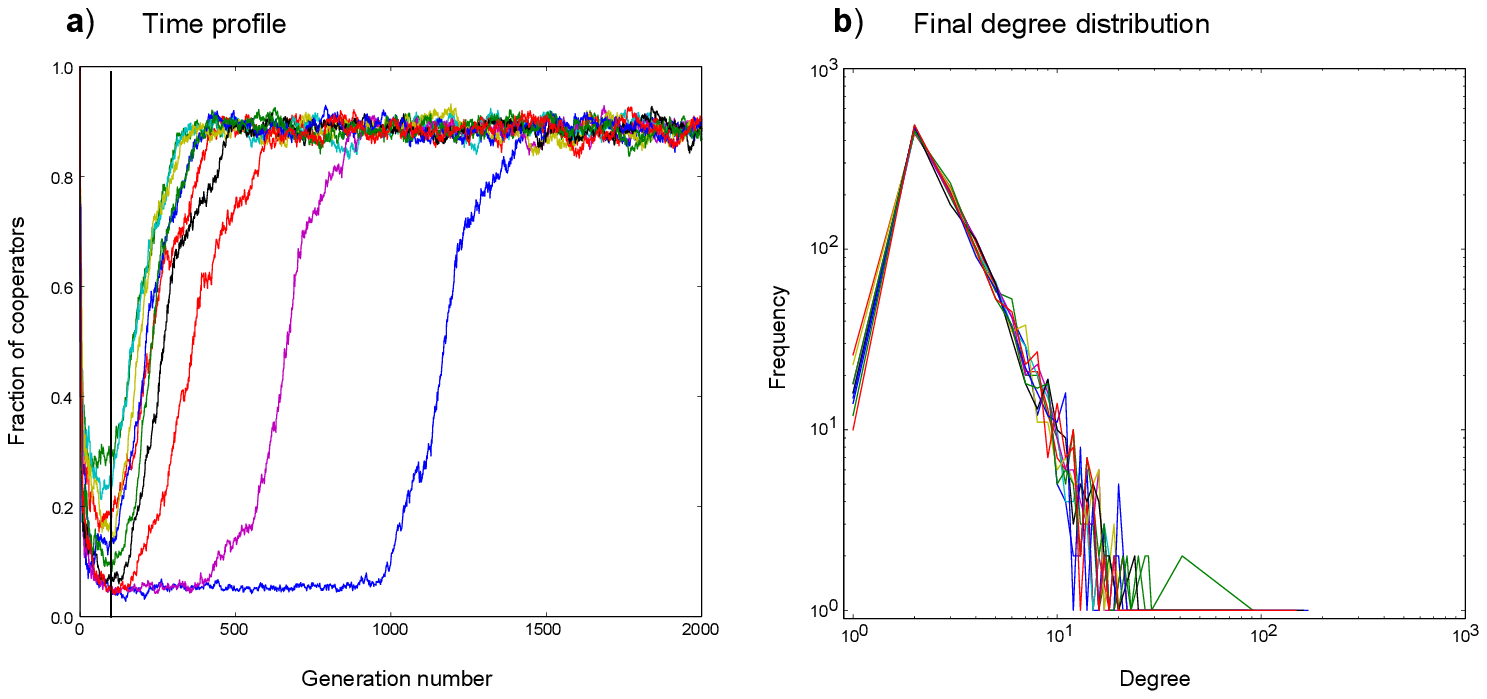}
	\caption{Time profiles and corresponding final degree distributions for networks grown from cooperator founders using the fluctuation model. (\textbf{a}) shows the time profile for a simulation consisting of 10 replicates with a $b$ value of 2.2. The fluctuation model is truncating networks using 2.5\% truncation. (\textbf{b}) shows the final degree distributions (at generation 2000) for each of the 10 simulation replicates. }
	\label{fig:Time_profiles_with_degree_distributions}
\end{figure}

Figure \ref{fig:Time_profiles_with_degree_distributions} illustrates time profile plots, with corresponding network degree distributions, for replicate simulations grown from cooperator founders using the fluctuation model. We see that the fluctuation model enables all replicates to consistently reach an \emph{apparent} power-law degree distribution, as previously reported for the EPA model\cite{poncela_growing_2012}{}. We also observe the same result (not shown) for the fluctuation model operating on networks grown from defector-founded populations. In addition, the replicate data makes clear that, when cooperation arises, variability in transition times (Figure \ref{fig:Time_profiles_with_degree_distributions}a) does not correspond to variability in degree distribution (Figure \ref{fig:Time_profiles_with_degree_distributions}b).

The presence of a small number of nodes with degree $k = 1$ is an artefact of the fluctuation model implementation. The fluctuation model grows the network in the same way as EPA (with each new node extending $m = 2$ connections), however the truncation component of the fluctuation model can leave residual nodes of degree $k = 1$ (at low frequencies) due to the deletion of connections from removed nodes. 

\begin{figure}[!]
	\centering
	\includegraphics[width=17cm]{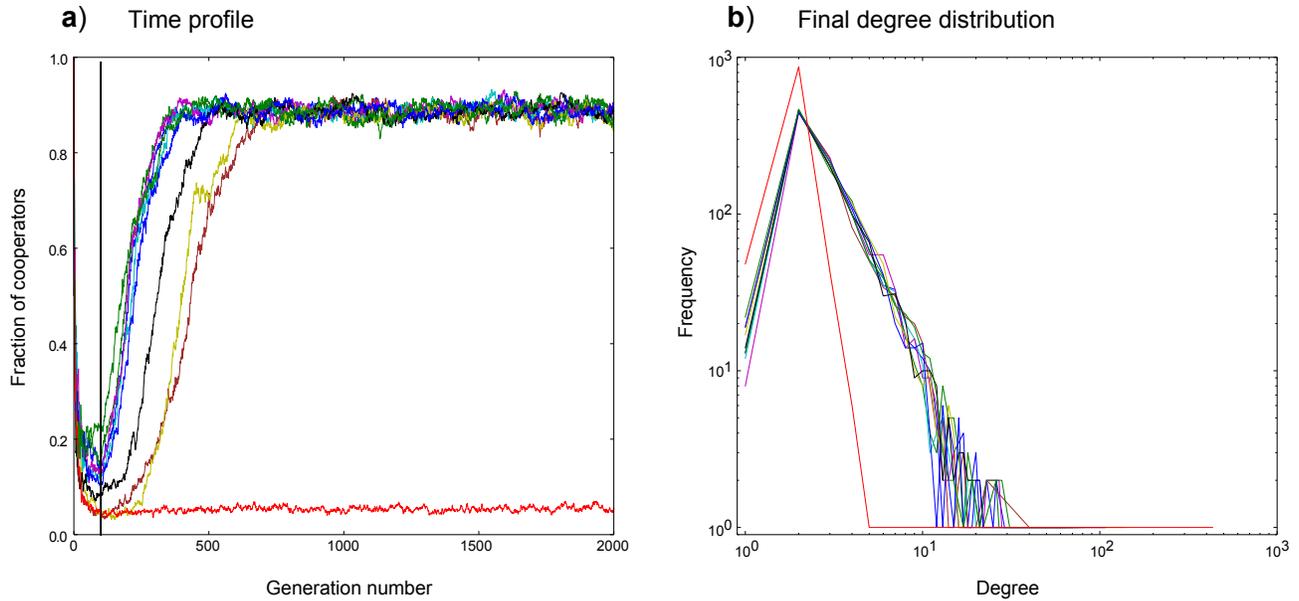}
	\caption{Plot illustrating the difference in degree distribution observed for a replicate that fails to achieve cooperation (fluctuation model).  Networks were grown from cooperator founders.  Simulation consists of 10 replicates with a $b$ value of 2.2. The fluctuation model is truncating networks using 2.5\% truncation. (\textbf{a}) shows the time profile. (\textbf{b}) shows the final degree distributions (at generation 2000) for each of the 10 replicates.  The red line in (\textbf{a}) (defectors predominate the population) corresponds to the red line in (\textbf{b}) (steeper exponent than all other replicates).}
	\label{fig:degree_dis_when_cooperation_fails}
\end{figure}

\noindent\textbf{Cooperation has a characteristic degree distribution.} Whilst in the majority of cases, the fluctuation model supported a transition of networks to a higher level of cooperation, we observed that as $b$ values increased, the transition was not guaranteed. Figure \ref{fig:degree_dis_when_cooperation_fails} captures an example of this, for 1 replicate out of 10 (for $b = 2.2$).  The replicate data demonstrates clearly the difference in degree distributions between networks that transition to cooperation and those that do not (the red lines in plots \ref{fig:degree_dis_when_cooperation_fails}a and \ref{fig:degree_dis_when_cooperation_fails}b refer to the same replicate).

\subsection*{Results for pre-existing random networks}
The following results look at the effect of the fluctuation model when applied to pre-existing random networks. \\

\begin{figure}[!]
	\centering
	\includegraphics[width=8.7cm]{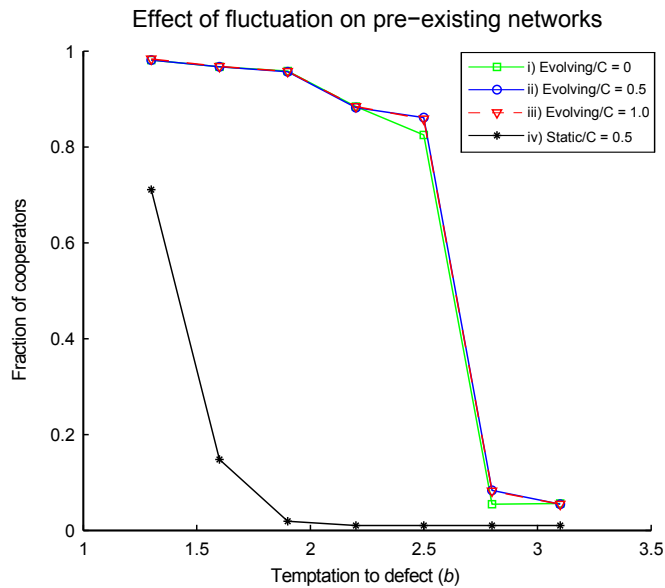}
	\caption{Effect of fluctuation model on pre-existing random networks. The plot shows temptation to defect plotted against final fraction of cooperators. Each data point represents an average of 25 individual results. Each of the individual results is an average of the last 20(of 20,000) generations of a simulation replicate. The fluctuation model used 2.5\% truncation. The pre-existing networks were in the form of random graphs  with each node in the network being populated by cooperators according to probabilities: \textbf{i}) 0, \textbf{ii}) 0.5 and \textbf{iii}) 1.  For reference, simulations involving a network that was structurally immutable are also shown in \textbf{iv}.  For the immutable network, nodes were populated with cooperators (or defectors) according to a probability of 0.5.}
	\label{fig:fluctuation_b_profiles_pre_existing_random_networks}
\end{figure}

\noindent\textbf{Fluctuation drives non-cooperative pre-existing networks to cooperation.} Figure \ref{fig:fluctuation_b_profiles_pre_existing_random_networks} shows final levels of cooperation achieved in simulations which started from randomly structured networks. Nodes within these networks were allocated cooperator (defector) strategies according to probability $P$ ($1-P$). Simulations were run for 20,000 generations during which time the majority ($>95\%$) of replicates transitioned to cooperation (for those simulations using $b$ values where cooperation was seen to emerge). Three pre-existing networks were tested, consisting of i) all cooperators, ii) cooperators and defectors in approximately equal amounts, and iii) all defectors. The curves for these three networks are almost entirely coincident, again illustrating the emergence of cooperation in the fluctuation model, regardless of starting criteria (as seen previously in networks grown from founder populations).  A static network (iv), where structural changes were disallowed (i.e. strategy updating only), is shown for comparison and clearly illustrates the contribution of the fluctuation mechanism.\\

\begin{figure}[!]
	\centering
	\includegraphics[width=17cm]{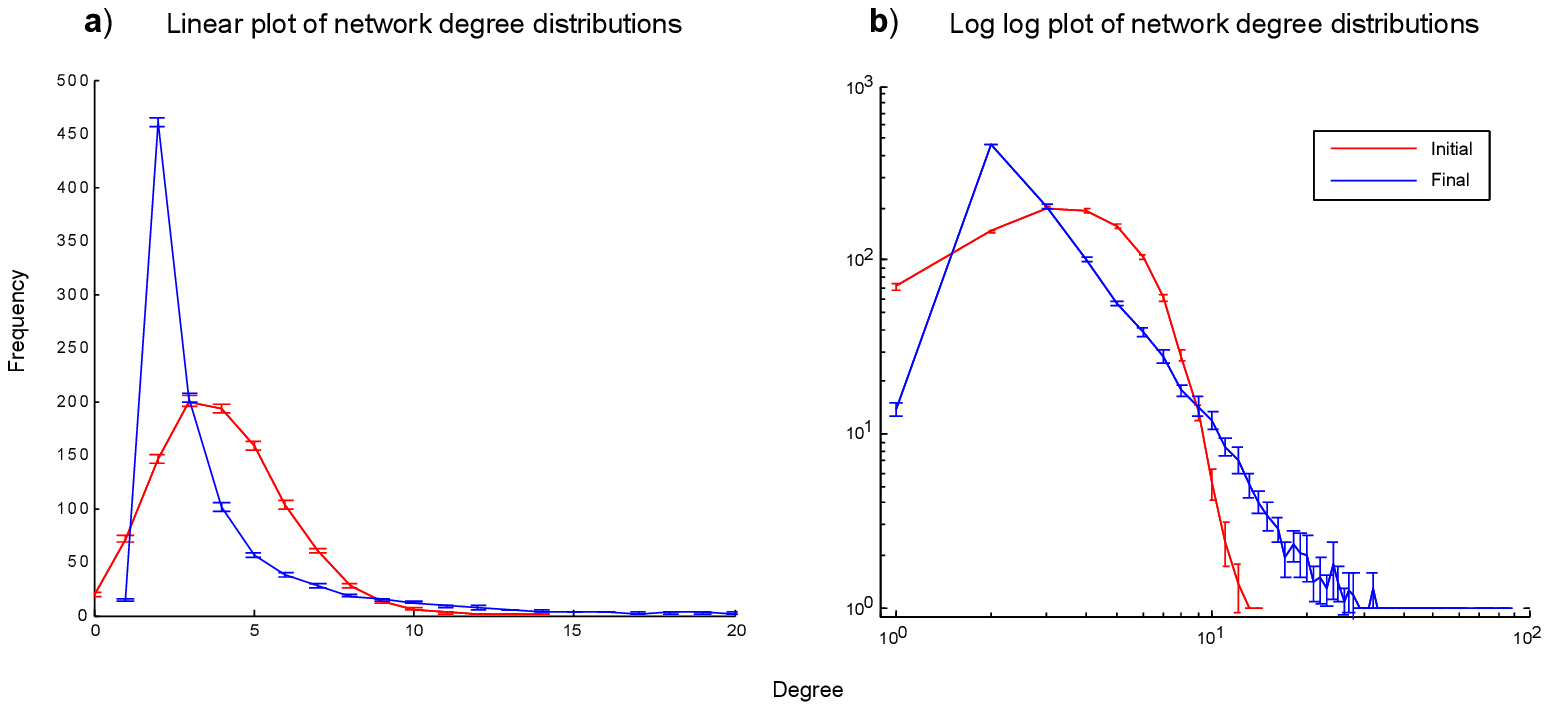}
	\caption{Degree distributions for pre-existing network at start and end of fluctuation model simulations. The plots present aggregate data from fluctuation model simulations of 25 replicates and illustrate the starting and finishing degree distributions, after 20,000 generations.  The simulations used a $b$ value of 2.2 and truncation at 2.5\%. The starting networks were in the form of random graphs populated entirely by defectors. The same data are represented on linear plots (\textbf{a}) and log log plots (\textbf{b}) in order to clearly illustrate, respectively, the apparent Poisson initial and power-law final distributions. In the interests of visualising both curves, the linear graph only includes degree values up to $k = 20$. The error bars shown represent 95\% confidence intervals for the data.} 
	\label{fig:Degree_distributions_pre_existing_network_start_and_end}
\end{figure}

\noindent\textbf{Fluctuation transforms pre-existing network structure from random to scale-free.} In Figure \ref{fig:Degree_distributions_pre_existing_network_start_and_end} we show the effect of the fluctuation model on degree distribution, for pre-existing random networks, initially composed entirely of defectors.  Figure \ref{fig:Degree_distributions_pre_existing_network_start_and_end}a, using linear axes,  highlights the initial Poisson degree distribution for the pre-existing random network, and Figure \ref{fig:Degree_distributions_pre_existing_network_start_and_end}b highlights apparent log-log linearity of the final degree distribution, that is characteristic of a power-law distribution.  \\

\noindent\textbf{Cooperation appears to be permanent.} In several thousands of simulations, excluding the small fluctuations visible in asymptotic states (see Figures \ref{fig:Example_simulation_plots_C_founders}, \ref{fig:Example_simulation_plots_D_founders} and \ref{fig:Time_profiles_with_degree_distributions}), whilst we have observed failures to transition to cooperation, we have not observed a single instance of widespread reversion to defection once cooperation has been achieved within a population. It would appear that once cooperation is established by means of this model, it does not collapse.\\

\section*{Discussion}
The main findings of our investigations are that:
\begin{enumerate}[label=\roman*, nolistsep]
	\item fluctuation of population size leads to an increase in levels of cooperation compared with the EPA model, 
	\item that the levels of cooperation achieved thus are largely independent of whether the populations were founded from defectors or cooperators,
	\item that the fluctuation model supports the emergence of cooperation both in networks grown from founder populations and also pre-existing random networks. 
\end{enumerate}

\medskip

The time profile plots we have provided in our results, give an indication as to how the fluctuation model is able to reach the increased levels of cooperation. Whilst the EPA model results in a high degree of variability, the fluctuation model produces consistent transition profiles. The EPA model has two interacting dynamic components: preferential attachment and strategy updating. Structural organization within the EPA model, which is driven by the preferential attachment component, occurs until the network reaches its defined size limit. Changes in levels of cooperation continue to  occur after this point. Given that the network structure is fixed, these latter changes can only occur as a result of the remaining active component of the EPA model process: strategy updating. Close examination of EPA simulation replicate time profiles (as shown in Figure \ref{fig:Example_simulation_plots_C_founders}) reveals an interesting observation (for $b$ values greater than 1.6): At the point where network structure becomes fixed, those replicates having higher levels of cooperation at this time, tend to finish with higher levels of cooperation than replicates experiencing lower levels of cooperation at the network fixation point. Whilst we do not yet have a detailed understanding of how cooperative structure develops within our networks, this observation suggests that prior to the network fixation point, some structural precedent is set which gives a probabilistic indication of how a network will profit, in cooperative terms, from strategy updating subsequent to structure fixation. 

Based on the work of Poncela et al.\cite{poncela_growing_2012} which describes the connection between scale-free network structure and cooperation, a plausible explanation for such a structural precedent is as follows: Whilst new nodes are preferentially attached to a growing network in a way that may generate hubs and hence a scale-free structure, there is no guarantee that the early clusters of nodes appearing in the network will be cooperators (cooperator and defector strategies are assigned to newly added nodes with equal probability). If the first network hubs appearing in the network are largely occupied by cooperators who in turn have cooperative neighbours, then these agents are likely to accumulate high fitness scores. This would potentially set the foundation for cooperation since such a group is likely to have a hub which would then draw further connections from newcomer nodes and hence promote scale-free structure. On the other hand, if early groupings of cooperators are interspersed with large numbers of defectors, this is likely to result in defectors preying on cooperators and initially accumulating higher fitness values. Strategy updating around such groups would in turn result in the conversion of cooperators present to the (at that point in time) more successful defectors. Eventually large groups of defectors will arise and result in lower fitness values. In a sea of low fitness values, preferential attachment is less able to demonstrate the ``rich get richer effect'' and this is more likely to result in random rather than targetted connections for newcomer nodes. After network fixation, strategies can be updated, but the network structure cannot change and early groupings of defectors in this way are likely to disrupt, impair or sufficiently delay the foundation of the scale-free structure that is needed to support higher levels of cooperation.  We anticipate forming a testable hypothesis around this explanation as the basis for a subsequent work on this model. 

The fluctuation model effectively allows a network to ``go back in time'' to fix structure that may have caused such a poor start. The model targets low fitness nodes (and their edges) for deletion. Such nodes are more likely to be defectors surrounded by other defectors (a defector–defector interaction results in a payoff of zero for both parties). Replacement nodes do not inherit the deleted node’s connections; they are preferentially attached to higher fitness nodes. In this way, networks that have a poor start are no longer permanently fixed, they have repeated opportunities to address the `poorest performing' elements of their structure. When viewed in this way, it is no longer surprising that similar levels of cooperation are ultimately achieved regardless of starting strategies. In the same way that this process of continual readjustment allows the network to deal with a less-than-ideal initial structure, it similarly allows the network to deal with less than favourable starting strategies.  If such strategies perform poorly, then sooner or later there is a likelihood they will be deleted, and should their replacements also perform poorly, there is a similar likelihood that they too will be deleted.

It is this ability to continually replace poor performing network nodes and connections that supports the fluctuation model’s striking ability to convert pre-existing random networks, initially populated entirely by defectors, to highly cooperative networks with a power-law distribution.

The fluctuation model studied in this paper is not intended as an accurate representation of any specific real world process, and probably does not map onto any such process precisely.  However it may be interpreted in several ways as analogues of natural phenomena. We now briefly consider possible interpretations of specific aspects of our model.  Firstly, as in the original EPA model, new nodes joining the network could be considered as being ``newly born into the network'' or as ``newcomers from outside the network''. In either case, they are positioned by preferential attachment in network `locations' which may prove advantageous or disadvantageous to them. Secondly, given that the model is one of (Darwinian) evolution, we tend to view strategy updates as equivalent to new population members replacing old, rather than any form of learning or imitation. This may be viewed as birth-death. In this situation, the new strategy `inherits' a set of connections forged by its ancestors along with the advantages or disadvantages that those connections confer. Thirdly, fluctuation as we have implemented it, deletes not only the least fit agents, but also the connections established over generations by successive offspring at that network location. The purpose of deleting both agent \emph{and} network node is to introduce some form of flux into the actual network itself rather than just its constituents' behaviour. This is a different and more disruptive process to that described by strategy updating - perhaps akin to real world scenarios where external environmental effects may have wider consequences for an entire population than for just specific individuals. 

Each of these processes is open to alternative interpretations. However, we suspect based on our results from this work, that it is not necessarily the exact process that is the important issue here, it is merely that, much like most ecological  systems in nature, a network continues to be perturbed in some way and is hence unable to achieve a permanently fixed structure: it thus continues to adapt. We anticipate that there may be alternative ways of perturbing a similar model to achieve results akin to those we have demonstrated.

\section*{Conclusion}
In summary, natural selection acts as a culling process that maintains diversity. In this work we have attempted to generalise the effect of that process across a model of networked individuals – with the crucial ability that individuals can locally affect their network in response to such culls. We have introduced a relatively simple modification to the original EPA model, symbolising an effect elemental to the behaviour of populations in the real world – effectively some sort of representation of flux in the environment.  This modification creates the opportunity for individuals in a population to continue to test adaptation against the selective pressures of the ecosystem. We have shown that such a feature brings about marked increases in levels of cooperation in networks grown from defector-founded populations. We have also shown that this feature results in levels of cooperation which are independent of starting behaviour and we have shown that the model can bring about cooperation in both growing and pre-existing non-cooperative networks. 

It is important that models which seek to explain the origins of cooperation are general and also robust to starting conditions. We believe that our model achieves both of these aims and hence our findings are of value in the search to understand the emergence and the ubiquity of cooperation.

\section*{Methods}
Our model and simulations are based on those described in \cite{poncela_complex_2008}{}, but with the addition of a pruning step which  deletes nodes from the network. We here give a full description of the approach for completeness.

\subsection*{Overview of approach}
The models consist of a network (i.e. graph) with agents situated at the nodes. Edges between nodes represent interactions between agents. Interactions are behaviours between agents playing the one-shot prisoner's dilemma game. These behaviours are encoded by a `strategy' variable, associated with each agent, which takes one of two values: cooperate or defect. The game is played in a round robin fashion, with each agent playing its strategy against all its connected neighbours, in turn. Each agent thus accumulates a fitness score which is the sum of all the individual game payoffs. 

Within an evolutionary simulation, starting from a founding population, this process is repeated over generations. The evolutionary process assesses agents at each generation on the basis of their fitness score; fitter agents' strategies remain unchanged; less fit agents are more likely to have strategies displaced by those of fitter neighbours. The evolutionary preferential attachment (EPA) process \cite{poncela_complex_2008} connects strategy dynamics to network growth: Starting from a small founding population new nodes are added which preferentially connect to fitter agents within the network. 

Our adaptation of the EPA model adds a further component which repeatedly truncates the network: Whenever the population reaches a maximum size, a specified percentage of nodes in the network are removed, on the basis of least fitness, after which the network grows again. 

\subsection*{Outline of the evolutionary process.} 
As described earlier, the general evolutionary process we implement is as follows:

\begin{enumerate}[noitemsep]
	\item \textit{Play prisoner's dilemma}
	\item \textit{Update strategies}
	\item \textit{Grow network}
	\item \textit{Remove nodes (only in the case of attrition models)}
\end{enumerate}

\noindent In the following, we provide more detail on the specifics of each of the four steps:\\

\begin{figure}[h]
	\begin{center}
		\includegraphics[width=6cm]{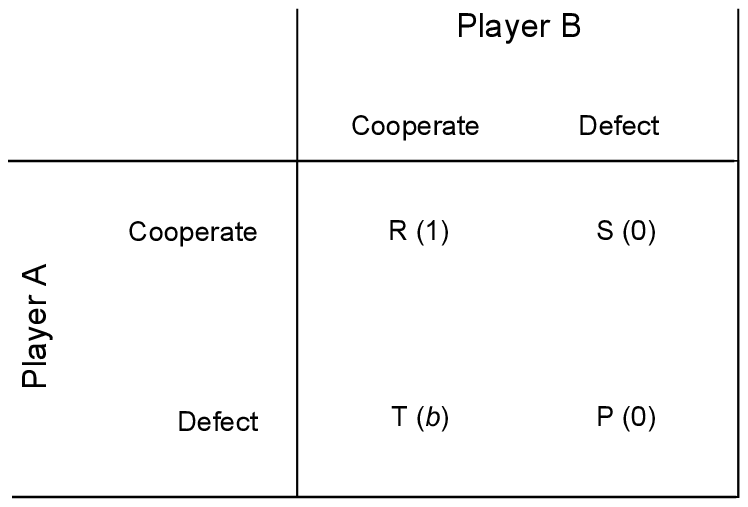}
		\caption{Payoff matrix for weak prisoner's dilemma.}
		\label{fig:PD_payoff_matrix}
	\end{center}
\end{figure}

\noindent \textbf{\textit{Play prisoner's dilemma}}. We use the single parameter representation of the one-shot prisoner's dilemma as formulated in \cite{nowak_evolutionary_1992}{}. In this form (the `\textit{weak}' prisoner's dilemma), payoff values for the actions, referred to as \textit{T, R, P} and \textit{S}, become $b$, 1, 0 and 0 (see Figure \ref{fig:PD_payoff_matrix}). The \textit{b} parameter represents the `temptation to defect' and is set at a value greater than 1 for the dilemma to exist. \\

From the accumulated prisoner's dilemma interactions, each agent achieves a fitness score as follows:

\begin{equation}
f_i = \sum_{j=1}^{k_i} \pi_{i,j},
\label{eqn:fitness_scoring}
\end{equation}
\noindent where $k_i$ is the number of neighbours that node \textit{i} has, \textit{j} represents a connected neighbour and $\pi_{i,j}$ represents the payoff achieved by node \textit{i} from playing prisoner's dilemma with node \textit{j}.\\

\noindent \textit{\textbf{Update strategies}}. Each node \textit{i} selects a neighbour \textit{j} at random. If the fitness of node \textit{i}, $f_i$ is greater or equal to the neighbour's fitness $f_j$, then \textit{i}'s strategy is unchanged. If the fitness of node \textit{i}, $f_i$ is less than the neighbour's fitness, $f_j$, then \textit{i}'s strategy is replaced by a copy of the neighbour \textit{j}'s strategy, according to a probability proportional to the difference between their fitness values. Thus poor scoring nodes have their strategies displaced by the strategies of more successful neighbours. 

Hence, at generation \textit{t}, if $f_{i}(t)\geq f_{j}(t)$ then \textit{i}'s strategy remains unchanged. If $f_{i}(t)< f_{j}(t)$ then \textit{i}'s strategy is replaced with that of the neighbour \textit{j} with the following probability:

\begin{equation}
	P_i = \frac
	{f_j(t) - f_i(t)}
	{b.max[k_i(t),k_j(t)]},
	\label{eqn:strategy_updating}
\end{equation}

\noindent where $k_{i}$ and $k_{j}$ are degrees of node \textit{i} and its neighbour \textit{j} respectively. The purpose of the denominator is to normalise the difference between the two nodes. The term $b.max[k_{i}(t),k_{j}(t)]$ represents the largest achievable fitness difference between the two nodes given their respective degrees. (The highest payoff value in the prisoner's dilemma is \textit{T}, equivalent to \textit{b} in the single-parameter version of the game used here. The maximum possible score for a node of degree \textit{k} is therefore \textit{$k*b$}. The lowest payoff value is \textit{P} or \textit{S}, both equal to zero, giving $k*b = 0$. Thus the maximum possible difference between two nodes is simply the maximum possible score of the fitter node.)\\

\noindent \textit{\textbf{Grow network}}. New nodes with randomly allocated strategies are added, to achieve a total of 10 at each generation. Each new node uses \textit{m} edges to connect to existing nodes. In all our simulations, we use \textit{m} = 2 edges. Duplicate edges and self-edges are not allowed. The probability that an existing node \textit{i} receives one of the \textit{m} new edges is as follows:

\begin{equation}
	\Pi(t) = \frac
	{1 - \epsilon + \epsilon f_i(t)}
	{\sum_{j=1}^{N(t)}(1 - \epsilon + \epsilon f_j(t))} ,
	\label{eqn:EPA_node_addition}
\end{equation}

\noindent where $f_i(t)$ is the fitness of an existing node \textit{i} and $N(t)$ is the number of nodes available to connect to at time \textit{t} in the existing population. Given that in our model each new node extends $m = 2$ new edges, and multiple edges are not allowed, $N$ is therefore determined \textit{without replacement}. The parameter $\epsilon \in [0,1)$ is used to adjust selection pressure. For all of our simulations $\epsilon = 0.99$, hence focusing our model on selection occurring directly as a result of the preferential attachment process.\\

\noindent \textit{\textbf{Truncate network}}. On achieving a specified size, the network is truncated according to a percentage $X$. Truncation is achieved by ranking all nodes in order of current fitness scores from maximum to minimum. The $X$ least fit nodes are then deleted from the network. All edges from deleted nodes are removed from the network. Any nodes that become disconnected from the network as a result of this process are also deleted. (Failure to do this would result in small numbers of single, disconnected, non-playing nodes, having static strategies, whose zero fitness values would result in continual isolation from the network.) When there are multiple nodes of equivalent low fitness value, the earliest (oldest) nodes are deleted first. Where $X$ = 0, no truncation occurs and the fluctuation model becomes the EPA model.

\subsection*{Investigations of the fluctuation model in networks grown from founder populations}
We investigated networks grown from an initial complete network with $N_0 = 3$ agents at generation $t_0$. Founding populations were either entirely cooperators or entirely defectors. We tested a range of different sized truncation values from 0.001 to 50\% starting from each of the two founder populations (cooperators or defectors). Networks were grown to a maximum size of $N = 1000$ nodes with an overall average degree of approximately $k = 4$. Simulations were run until 2000 generations. The `fraction of cooperators' values we use are means, averaged over the last 20 generations of each simulation (to compensate for variability that might occur if just using final generation values). Each data point on `b-profile' plots (Figures \ref{fig:Effect_of_truncation}, \ref{fig:Effect_of_random_selection_for_truncation}, and \ref{fig:fluctuation_b_profiles_pre_existing_random_networks}) is the mean of 25 simulations. Simulations run for the purposes of illustrating time profiles or degree distributions were limited to 10 replicates in the interests of clarity. 

\subsection*{Investigations of the fluctuation model applied to pre-existing random networks}
Random networks were generated by randomly connecting edges between a specified number of nodes (i.e. maximum size of network).  Total number of edges added $N*k/2$, was determined on the basis of a random graph of degree $k=4$. Simulation parameters were as described previously for founder population investigations except, i) we focused on a single truncation value of $X = 2.5\%$ and ii) longer run times (e.g. 20,000 generations) were generally required for replicate simulations to stabilise, when looking at pre-existing networks initially populated entirely with defectors.

In applying the fluctuation model to pre-existing networks, the model simply `sees' a pre-existing network, as a 'grown-from-founders' network which has reached the point where it requires truncation. In essence, the fluctuation model is the same when it is applied to pre-existing networks as it is when applied to networks grown from founders.\\

\noindent Where parameters were modified from those described in this methods section (e.g. longer simulations), this is made clear in the results.

\bibliography{SGMreferencesYr3}

\section*{Acknowledgements}
This work was supported by funding from the Engineering and Physical Sciences Research Council (Grant reference number EP/I028099/1). 

\section*{Author contributions statement}
S.M. conceived and conducted the experiment(s), and analysed the results. J.K performed statistical analysis. Both authors reviewed the manuscript. 

\section*{Additional information}
The authors declare no competing financial interests

\end{document}